\documentclass[12pt]{article}
\usepackage{cite}
\usepackage{graphicx}
\pdfoutput=1

\begin{document}

\newcommand{\titulo}{Superconductivity in Nanosystems: A Fruitful Path to New Phenomenology in Quantum Materials}

\newcommand{\autor}{M.V.~Ramallo$^{a,b}$}

\newcommand{\direccion}{$^{a}$Quantum Materials and Photonics Research Group (QMatterPhotonics),\\ Department of Particle Physics, University of Santiago de Compostela,\\ 15782 Santiago de Compostela, Spain \\
$^{b}$Instituto de Materiais (iMATUS),\\ University of Santiago de Compostela, 15782 Santiago de Compostela, Spain}

\begin{center}
  \Large\bf
\titulo\\  \end{center}\mbox{}\vspace{-1cm}\\ 

\begin{center}\normalsize\autor\end{center} 

\begin{center}\normalsize\it\direccion\end{center}

\newcommand{\eg}{{e.g.}}
\newcommand{\ie}{{i.e.}}

\newcommand{\etal}{{et al.}}


\mbox{}\vskip0.5cm{\bf Abstract: }

We reason  that  various  recent works by different groups reporting new phenomenologies in superconductors can be understood in a unifying way as instances of the appearance of novel  competitions (or synergies in some cases) between the  coexisting orders at play in superconducting materials.   In particular, we argue that the main common feature of  such phenomenologies is to have emerged from the induction, by nanoengineering, of  novel  characteristic lengths for each order, or of custom regular spatial  patterns affecting them. We claim, thus, that a  fruitful path to discover new phenomenology is opened by these and future searches of novel nanostructurations of  superconducting materials.

\vspace{4cm}

\mbox{}\hfill{\footnotesize {\tt mv.ramallo@usc.es}}
\thispagestyle{empty}

\newpage
\setlength{\baselineskip}{18pt}
\



In the recent years, the landscape of the research in superconductivity has experienced a progressive focus on varied  superconducting systems that share as common primary {characteristics} the reduction of some of their dimensionalities and the emergence of qualitatively novel phenomenology with respect to bulk superconducting materials. 

This includes nanosystems, such as, \eg, very thin films~\cite{films},  tunnel--junction-based hybrid devices~\cite{tuneles}, superconductors with nanospaced defects or vortex pinning centers~\cite{pinning}, interfacial superconductivity~\cite{interfacial}, etc.

Superconductivity is a collective phenomenon in which spatial correlations and coherences  {are crucial}. 
 It is not surprising, then, that the effects of reduced dimensionality may affect the superconducting properties. Correspondingly, the nanosizing and/or nanostructuring of superconductors may lead to the advent of novel phenomenology not present in bulks and which is also qualitatively different from the one in other nanosystems due to the unique quantum properties of superconductors. The result is that the study of superconductivity in nanosystems is  uncovering rich, singular, and qualitatively novel phenomenologies, often quantum in character. The interest of these discoveries is two-fold. On the one hand, their study is academically sound per se. On the other hand, their study is relevant to various applications where superconductors are preferred materials. These phenomena are exemplified in the fields of quantum computing and communications, high magnetic field applications, quantum sensorics, etc.~\cite{nanoSC}

Examples abound of the growing interest in the interlinks between nanoscience and superconductivity. Here we will provide examples of recent works by different groups reporting new phenomenologies in superconductors, and  claim  that they can be understood in a unifying way as instances of the possibilities of employing nanoscience to  customize either the competitions or the synergies between the different coexisting orders at play in superconducting materials.

In the following paragraphs, we  briefly present three groups of example researches that have achieved the emergence of novel phenomena by means of nanoengineering-induced effects, and identify in them the relevant order competitions or synergies.

Firstly, the superconductivity  may be affected directly by the interplay of the intrinsic superconducting quantities (superconducting wave function, quantized vortices, etc.) and an additional, nanoengineered order. The first has a tendency to organize itself following characteristic lengths, such as the superconducting coherence length or the distance between vortices. Those lengths can be targeted by nanostructuring the coexisting order. 

For instance, let us mention here, as examples, some works dedicated to the interesting interrelations between magnetic and superconducting orders that become realizable through nanoscience.  Winarsih \etal~\cite{Win} conducted studies of nanoparticles of the \mbox{La$_{2-x}$Sr$_x$CuO$_4$} cuprate superconductor, observing that a reduction of the particle size down to the nanometric scale is accompanied by the appearance of magnetic correlation in the Cu spin fluctuations. They likewise observed a change in the bond distance between Cu and O ions in the conducting layer, which seems to correlate with that new magnetism.

Also, Vettoliere \etal~\cite{Tafuri} focused their attention on the case of Josephson tunnel nanojunctions, whereto a ferromagnetic element is added. They report the fabrication of such systems using multistage deposition--lithography processes. The so-obtained junctions become switchable using the influence of the magnetic element over the superconducting wave function, which may be of great relevance in the area of the emerging quantum computing and communications technologies~\cite{Tafuri,nanoSC}. The junctions are Al-based and bear good prospects of integrability in different types of quantum circuits.

Both of those works exemplify that nanoengineered regular implantation of magnetic order may induce novel superconducting phenomenology as a byproduct. 

Secondly, novel phenomenology has been also achieved  by producing an  increase, due to nanoengineering, of the action of pinning centers over the superconducting quantum vortices, that have their own independent ordering with phenomenological consequences. It is today well known that such pinning may increase the critical current, $j_c$, which is of crucial importance, \eg, for high-field applications~\cite{pinning,Aichner,nanoSC}. In their article, Ivan \etal~\cite{Crisan} conduct studies to demonstrate, among other aspects, that the frequency-dependent $j_c$ of their samples with self-assembled defects can be explained by using a new and simple, but effective, method involving AC susceptibility measurements. It is a useful development that also points to implications concerning the dependence of the pinning potential with the probing~current.

The article by Backmeister \etal~\cite{Lang} is another deft example of the richness of new possible phenomenologies obtainable through nanostructured pinning. This research reports the observation of an `ordered Bose glass state' for the superconducting vortex lines. For that, a superconducting system with a regular and hexagonal array of pinning centers is manufactured, with only $30$~nm spacing between them, which allows for the entanglement of vortices.

Finally, let us also mention articles that exemplify how new phenomenology can appear from the frustration of the characteristic lengths of the superconducting wave function by the finite sizes of the system. In particular, Adhikari \etal~\cite{Adh} measured the resistive superconducting transition in Fe-implanted NbN nanometric-thin films, observing the appearance of up to six different regimes in the vicinity of that transition. They managed to interpret this rich phenomenology in terms of a finite-size granular model of the material, and the corresponding considerations on the bosonic and fermionic electrical conduction channels {that are}  expected to coexist near the transition. The different regimes appear as a result of the progressive formation of local conduction clusters, as well as eventually as a result of percolation paths between them.

Also, the  article by Botana \etal~\cite{Botana} explores the precursor superconductor state as the resistive transition is approached, in this case in very thin films of cuprate superconductors composed of only a few layers of unit cells. Measurements in such films  are not satisfactorily reproduced by  the  critical phenomena equations for bulk samples, nor for single two-dimensional plane superconductors~\cite{Cieplak,Vina}. Botana \etal~\cite{Botana} presented model calculations for the precursor superconducting wave funtion, which are specifically adapted to the few-layer case, obtaining good agreement with the data.

Of course, these selected examples  cannot fully cover the entire broad spectrum of researches being currently done on the frontiers between superconductivity and nanoscience. Hopefully, however, they will serve both as a way to demonstrate that  nanosystems open a fruitful path to new superconducting phenomenology, and to show that the  here signaled order competitions may serve as  target  vectors to find  new effects that are certainly still waiting to be discovered by exploring different forms of  nanoscience actuations over superconducting systems.

\vspace{6pt}

{\bf Acknowledgements:} This work was supported by the Agencia Estatal de Investigaci\'on (AEI) and Fondo Europeo de Desarrollo Regional (FEDER) through project PID2019-104296GB-100 by Xunta de Galicia (grant GRC number ED431C 2018/11) and iMATUS (2021 internal project RL3).

\end{document}